\documentclass[aps,preprint,showpacs,floats,epsf,epsfig,nofootinbib,12pt]{revtex4}
\usepackage{graphicx}% Include figure files
\usepackage{epsfig}
\usepackage[dvips]{color}

\def\be{\begin{eqnarray}}
\def\en{\end{eqnarray}}
\def\non{\nonumber}
\def\la{\langle}
\def\ra{\rangle}
\def\pp{{\prime\prime}}

\def\vp{\varepsilon}

\begin{document}

\title{Transitions of $B_c\rightarrow \psi(1S,2S)$  and the modified harmonic oscillator wave function in LFQM  }

\vspace{1cm}

\author{ Hong-Wei Ke$^{1}$   \footnote{khw020056@hotmail.com}, Tan Liu$^{1}$ and
        Xue-Qian Li$^2$\footnote{lixq@nankai.edu.cn}  }

\affiliation{  $^{1}$ School of Science, Tianjin University, Tianjin 300072, China \\
  $^{2}$ School of Physics, Nankai University, Tianjin 300071, China }

\vspace{12cm}

\begin{abstract}
The LHC$_b$ collaboration has systematically measured the rates of
$B_c\to J/\psi K$, $B_c\to J/\psi D_s$, $B_c\to J/\psi D_s^*$ and
$B_c\to \psi(2S) \pi$. The new data enable us to study relevant
theoretical models and further determine the model parameters. In
this work, We calculate the form factors for the transitions
$B_c\to J/\psi$ and $B_c\to \psi(2S)$ numerically, then determine
the partial widths of the semi-leptonic and non-leptonic decays.
The theoretical predictions on the ratios of  $\Gamma(B_c\to
J/\psi K)/\Gamma(B_c\to J/\psi \pi)$, $\Gamma(B_c\to J/\psi
D_s)/\Gamma(B_c\to J/\psi \pi)$ and $\Gamma(B_c\to J/\psi
D_s^*)/\Gamma(B_c\to J/\psi \pi)$ are consistent with data within
only 1$\sigma$.  Especially, for calculating $\Gamma(B_c\to
\psi(2S)X)$ the  modified harmonic oscillator wave function (HOWF)
which we developed in early works is employed and the results
indicate that the modified  harmonic oscillator wave function
works  better than the traditional HOWF.


\end{abstract}

\maketitle

\section{Introduction}
Unlike charmonia and bottomonia which have been thoroughly
investigated from both experimental and theoretical aspects, the
researches on $B_c$ and its excited states are far behind  because
of lack of necessary data for a long while. The luminosity of LEP
I and II was not sufficient to produce $B_c$ \cite{Chang} and
$B_c-$the ground state of the meson family which contains two
different heavy flavors was eventually observed at hadron
colliders. Until now $B_c$ is still only measured at TEVATRON and
LHC. In the future proposed $Z_0$ and/or Higgs factories or even
ILC with very high luminosity will produce a
 large database of $B_c$ and their excite states
which can provide more accurate information about the two-heavy-flavor measons.

Charmonia and bottomonia mainly decay via strong and
electromagnetic interactions, instead $B_c$ can decay only via
weak interaction, therefore its lifetime is much longer than the
quarkonia. Even though LHC is a hadron collider, the background is
much messier than at electron-positron colliders, because of its
high energy and luminosity, LHC offers us an opportunity to study
$B_c$ and its excited states. Recently LHC$_b$ collaboration has
measured several decay modes of $B_c$ and obtained
$\Gamma(B_c\to\psi(2S)\pi)/\Gamma(B_c\to
J/\psi\pi)=0.25\pm0.068\pm0.014\pm0.006$\cite{Aaij:2013oya};
$\Gamma(B_c\to J/\psi K)/\Gamma(B_c\to
J/\psi\pi)=0.069\pm0.019\pm0.005$\cite{Aaij:2013vcx};
$\Gamma(B_c\to J/\psi D_s)/\Gamma(B_c\to
J/\psi\pi)=2.9\pm0.57\pm0.24$ and $\Gamma(B_c\to J/\psi
{D_s}^*)/\Gamma(B_c\to J/\psi
D_s)=2.37\pm0.56\pm0.10$\cite{Aaij:2013gia}. It would be a good
time to carry out serious theoretical studies on those decay modes
which may provide us more information about the structure of such
two-heavy-flavor mesons and especially serve as a probe for our
models which deal with the non-perturbative QCD. Though  the
typical $P\to V$ ($P$ and $V$ denote a pseudoscalar  meson and a
vector meson respectively) transitions have been studied by
various
approaches\cite{Cheng:2011qh,Bhattacharya:2008ke,Ali:2007ff,Li:2006jv},
the theoretical predictions on $B_c$ are few. In
Ref.\cite{Cheng:2003sm} Cheng et al. studied $P\to V$ transitions
in the light front quark  model (LFQM) which has been established
and applied by many researchers later
\cite{Jaus,Ji:1992yf,Cheng:2004cc,Cheng:1996if,Cheng:2003sm,Choi:2007se,Hwang:2006cua,Ke:2007tg,Ke:2009ed,Li:2010bb,Ke:2013zs}.
In this work we will apply the formula derived by Cheng et al. in
Ref.\cite{Cheng:2003sm} to study the semi-leptonic decay $B_c\to
J/\psi(\psi(2S)) e\bar\nu_e$ and non-leptonic decay $B_c\to
J/\psi(\psi(2S))+X$ ($X$ can be $\pi, K, K^*, D, D^*, D_s$ and
$D_s^*$). Hopefully we can further test the validity degree of the
LFQM and constrain the model parameter space.

In LFQM a phenomenological wave function  is introduced to describe  the  momentum
distribution amplitudes of the constituent quarks and the harmonic oscillator wave functions may be the most convenient and applicable one among all possible forms.
Most of the previous studies only explored the transitions between ground states.
In our early work \cite{Ke:2010x} we calculated the decay constants of $\Upsilon(nS)\;(n> 1)$ (excited states of bottomonia) with the traditional harmonic oscillator wave functions and found that
the theoretical results obviously conflict with data, so we proposed to choose a
modified harmonic oscillator wave function instead for the radially excited states. With this change, the inconsistency between theoretical predictions and data disappears.
In this work we would like to further check the modified harmonic oscillator wave functions for radially excited state in $B_c\to \psi(2S)$ weak decays. Comparing the theoretical results with data one can judge whether the
modified harmonic oscillator wave functions work better than the traditional ones.

After the introduction we present the relevant formulas  for $P\to V$ transition in section II where we introduce briefly our modified
harmonic oscillator wavefuncions. Then we
numerically evaluate the form factors and the decay widths
for the available decay modes and predict the rates for some channels which have not been measured yet. In the section, we also discuss
the results obtained in this theoretical framework. At last we make a brief summary.

\section{formulas}
\subsection{$P\rightarrow V$ transition in the LFQM}
The form factors for $B_c\rightarrow J/\psi$ and $B_c\rightarrow
\psi(2S)$  which are the typical $P\rightarrow V$ transitions are
defined as
 \be \label{eq1}
  \la V(p'',\vp^\pp)|V_\mu|P(p^\prime)\ra &=&
i\Bigg\{(M'+M'') \vp''^*_\mu A^{PV}_1(q^2)  - {\vp''^*\cdot
p'\over M'+M''}p_\mu A^{PV}_2(q^2) \non \\
&-& 2M'' {\vp''^*\cdot p'\over
q^2}q_\mu\left[A^{PV}_3(q^2)-A^{PV}_0(q^2)\right]\Bigg\},
\non \\
   \la V(p^\pp,\vp^\pp)|A_\mu|P(p^\prime)\ra &=& -{1\over
  M'+M''}\,\epsilon''_{\mu\nu\rho\sigma}\vp^{*\nu}p^\rho
  q^{\sigma}V^{PV}(q^2),
 \en
with
 \be A^{PV}_3(q^2)=\,{M'+M''\over 2M''}\,A^{PV}_1(q^2)-{M'-M''\over
2M''}\,A^{PV}_2(q^2),
 \en
where $M' (M'')$  and $p' (p'')$ are the mass and momentum  of the
vector  (pseudoscalar) state. We also define $p=p'+p''$ and
$q=p'-p''$.

As discussed in Ref.\cite{Cheng:2003sm} these form factors are calculated in the
space-like region with $q^+=0$, thus to obtain the physical amplitudes an extension to the time-like
region is needed.
To make the extension one may write out an analytical expressions for the form factors, and in
Ref.\cite{Cheng:2003sm} a three-parameter form was suggested
\begin{eqnarray}\label{eq2}
 F(q^2)=\frac{F(0)}{
  \left[1-a\left(\frac{q^2}{M_{\Lambda_b}^2}\right)
  +b\left(\frac{q^2}{M_{\Lambda_b}^2}\right)^2\right]}.
 \end{eqnarray}

\begin{figure}
\begin{center}
\begin{tabular}{ccc}
\scalebox{0.8}{\includegraphics{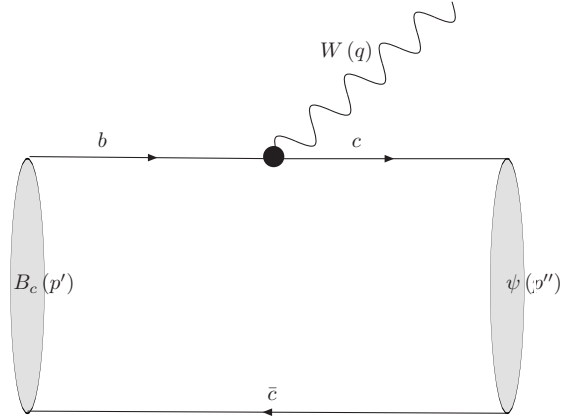}}
\end{tabular}
\end{center}
\caption{$B_c\rightarrow \psi$ transition}\label{fig1}
\end{figure}

The relevant Feynman diagrams for the transitions are shown in Fig.\ref{fig1}. In Ref.\cite{Cheng:2003sm} the
authors deduce all the detailed expressions for the form factors $A_0$, $A_1$, $A_2$ and $V$ in the covariant LFQM. One can
refer to Eq.(32) and (B4) of Ref.\cite{Cheng:2003sm} to find their explicit expressions.

\subsection{The modified harmonic oscillator wave functions}
For calculating the form factors $A_0$, $A_1$, $A_2$ and $V$,  the
light-front momentum distribution amplitudes need to be specified.
In most of such works, the harmonic oscillator wave function is
employed because of its obvious advantages. In our previous
work\cite{Ke:2010x} we found that predictions on the rates of the
processes where radially excited states are involved do not
coincides with data  as long as  the transitional harmonic
oscillator wave function was employed, thus we suggested to use a
modified harmonic oscillator wave function to replace the
traditional one for the radially excited states. It is found that
the modified wave function indeed works well when we calculate the
radiative decays of $\Upsilon(nS)\;(n>1)$.

The decay of $B_c\rightarrow \psi(2S)$  where $\psi(2S)$ is a
radially excited state would serve as an alternative probe for
testing the modified wave function.  Thus we use both the
traditional and modified harmonic oscillator wave functions to
calculate the rates of $B_c\rightarrow \psi(2S)+X$ where $X$
denotes some relevant mesons.  Through comparing the results
obtained in terms of the two kinds of $\psi(2S)$ wave function
with data, we can determine  their reasonability. The relevant
modified wave function is
\begin{eqnarray}\label{eq3}
\phi(1S)&&=4\Big(\frac{\pi}{\beta^2}\Big)^{3/4}\sqrt{\frac{\partial
k_z}{\partial x}}{\exp}\Big(-\frac{k^2_z+k^2_\perp}{2\beta^2}\Big),\nonumber\\
\phi(2S)&&=4\Big(\frac{\pi}{\beta^2}\Big)^{3/4}\sqrt{\frac{\partial
k_z}{\partial
x}}{\exp}\Big(-\frac{1}{2}\frac{k^2_z+k^2_\perp}{\beta^2}\Big)
\Big(3 -2\frac{k^2_z+k^2_\perp}{\beta^2}\Big),\nonumber\\
\phi_{\rm_M}(2S)&&=4\Big(\frac{\pi}{\beta^2}\Big)^{3/4}\sqrt{\frac{\partial
k_z}{\partial
x}}{\exp}\Big(-\frac{{2}^\delta}{2}\frac{k^2_z+k^2_\perp}{\beta^2}\Big)
\Big(a -b\frac{k^2_z+k^2_\perp}{\beta^2}\Big),
 \end{eqnarray}
where $\beta$ is a phenomenological  parameter and needs to be
fixed by fitting data. $k$ is the relative momentum of the constituents
and $x$ is the momentum fraction of the quark while $1-x$ is for
the anti-quark. More details can be found in
Ref.\cite{Ke:2010x,Cheng:2003sm}. In Ref.\cite{Ke:2010x} we fixed
$a=1.89$, $b=1.55$, $\delta=1/1.82$ for $\Upsilon(2S)$ and by the
heavy quark effective theory it is reasonable to suppose that they
are the same for $\psi(2S)$.

\subsection{ Rates of the semi-leptonic and non-leptonic decays }
Since no strong interaction in
the final states to contaminate the processes, semi-leptonic decays can shed more light for understanding the meson structure which is associated with non-perturbative QCD and
help to fix the model parameters. The amplitude for
the semi-leptonic decay is
 \be\label{eq4}
 \la \psi l \bar\nu_l | {\cal H} | B_c\ra=
 \frac{G_F}{\sqrt 2}V_{cb}
 \la V | V_\mu-A_\mu | P\ra \bar l \gamma^\mu(1-\gamma_5)\nu_l.
 \en

For evaluating the rates of non-leptonic decays $P \to V+ X$,
generally factorization is assumed i.e. the hadronic transition
matrix element can be factorized into a product of two independent
matrix elements: the transition matrix $<P|J'_{\mu}|V>$ and
$<0|J^{\mu}|X>$ which is determined by a decay constant.
For the non-leptonic decays $B_c \to
J/\psi(\psi(2S)) X$ the effective interaction at the quark level
$b\to c\bar{q_1}q_2$  is
 \be\label{eq5}
 &&{\cal H}_W=\frac{G_F}{\sqrt 2}V_{cb}V_{q_1q_2}^*(c_1O_1+c_2O_2),
 \en
where $c_i$ denote the Wilson coefficients and  $O_i$ are
four-quark operators. The hadronic transition matrix elements is
 \be\label{eq6}
 \la \psi M | {\cal H}_W | B_c\ra &&=
 \frac{G_F}{\sqrt 2}V_{cb}V_{q_1q_2}^* a_1\la V | V_\mu-A_\mu |P\ra
  f_{_M} q^{\mu}\,\,\,\, M {\rm \,is\, a \,pseudoscalar},
  \\&&=
 \frac{G_F}{\sqrt 2}V_{cb}V_{q_1q_2}^* a_1\la V | V_\mu-A_\mu |P\ra
  m_{_M}f_{_M} \varepsilon_{_M}^{\mu}\,\,\,\, M {\rm \,is\, a \,vector},
 \en
where  the Wilson coefficient $a_1=c_1+c_2/N_c$ with $N_c$ being
an effective color number which is 3 when the color-octet
contributions are not taken into account\cite{Buras:1985yx}.

\section{numerical results}
In this section we will calculate the form factors for these $P\to
V$ transitions. The masses $m_{B_c}=6.277$ GeV, $m_{J/\psi}=3.096$
GeV and $m_{\psi(2S)}=3.686$ GeV are taken from the Data-book
\cite{PDG12}. The parameter $\beta$ in the wavefunction of
$J/\psi$ is fixed to be 0.631 GeV when $m_c=1.4$
GeV\cite{Ke:2011jf}. However until now there are no available data
to fix the model parameter $\beta$ in the wavefunction of $B_c$,
so we will make an estimate based on reasonable arguments. In
Ref.\cite{Ke:2010x} we fixed $\beta=1.257$ GeV  for $\Upsilon$
where $m_b=5.2$ GeV was set, accordingly we take an average of
0.631 GeV and 1.257 GeV as the value of $\beta$ in the
wavefunction of $B_c$ which is fixed to be 0.944 GeV. In our
calculation we set $m_c=1.4$ GeV and $m_b=5.2$ GeV. The CKM matrix
elements take values: $V_{bc}=0.0406$,  $V_{cd}=0.2252$,
$V_{ud}=0.97425$, $V_{us}=0.2252$ and $V_{cs}=1.006$\cite{PDG12}.
The decay constants and masses for the relevant mesons are listed
in table \ref{Tab:t1}.
\begin{table}
\caption{Meson decay constants  and masses (in units of
 MeV).}
\begin{ruledtabular}\label{Tab:t1}
\begin{tabular}{cccccccc}
  meson & $\pi$  & $K$ & $K^*$ & $D$ & $D^*$ & $D_s$ & $D_s^*$  \\\hline
  $m$\cite{PDG12}& 139.6&493.7&891.7&1869.6&2010.3&1968.5&2112.3\\
  $f$\cite{Cheng:2003sm}   & 131      & 160 & 210   & 200 & 220   & 230   & 230
\end{tabular}
\end{ruledtabular}
\end{table}

%\begin{table}[!h]
%\caption{The  form factors given in the
%  three-parameter form ( $\beta=0.631$GeV$^{-1}$).}\label{Tab:t1}
%\begin{ruledtabular}
%\begin{tabular}{cccc|cccc|cccc}
%  $F^{B_cJ/\psi}$    &  $F(0)$ &  $a$  &  $b$ & $F^{B_c\psi(2S)}$& $F(0)$ &$a$&$b$& $F^{B_c\psi_M(2S)}$& $F(0)$&  $a$  &  $b$\\\hline
%  $A_{0}$  &   -4.43      &   5.38    &  8.78&7.39&2.48& $g_{12}$&-1.24 &  2.23  &  2.98&2.30&0.75  \\
%  $A_{1}$  &  -0.00204     &  1.62    & 1.38&0.65&0.13& $f_{22}$& 0.0204 &  2.43  & 3.34&2.60&0.847   \\
%  $A_{2}$  &    -4.08    &     -0.588  &  -2.13&-2.06&-0.724& $g_{32}$& $-0.396$ &  3.40  &5.15&4.13&1.36  \\
%  $V$  &      -0.145   &     1.21  & 1.05&0.612&0.168& $f_{42}$& $-0.0263$ &  2.14  &
%  2.79&2.12&0.68
%\end{tabular}
%\end{ruledtabular}
%\end{table}

\begin{table}[!h]
\caption{The  form factors given in the
  three-parameter form.}\label{Tab:t2}
\begin{ruledtabular}
\begin{tabular}{cccc|cccc}
  $F$    &  $F(0)$ &  $a$  &  $b$   &$F$    &  $F(0)$ &  $a$  &  $b$\\\hline
  $A^{B_c J/\psi}_{0}$  &   0.502      &   1.66    &  2.04  & $A^{B_c J/\psi}_{1}$  &  0.467     &  1.51    & 0.95 \\
   $A^{B_c J/\psi}_{2}$  &    0.398    &     1.97  &  1.84 &  $V^{B_c J/\psi}$  &      0.638   &     2.15  & 2.21\\
    $A^{B_c \psi(2S)}_{0}$  &   0.452      &   0.92    &  0.50  &  $A^{B_c \psi(2S)}_{1}$  &  0.335     &  -0.21    & 0.88  \\
    $A^{B_c \psi(2S)}_{2}$  &    0.102    &     -2.73  &  4.63 &  $V^{B_c \psi(2S)}$  &      0.525   &     0.53  & 0.96\\
 $A^{B_c \psi_M(2S)}_{0}$  &   0.300      &   1.15   &  0.60  &  $A^{B_c \psi_M(2S)}_{1}$  &  0.251     &  -0.058    & 0.98 \\
 $A^{B_c \psi_M(2S)}_{2}$  &    0.109    &     -1.93  &  3.71 &  $V^{B_c \psi_M(2S)}$  &      0.388   &     0.68  & 1.16
\end{tabular}
\end{ruledtabular}
\end{table}

With these parameters we calculate the form factors for the transitions $B_c\to J/\psi$ and $B_c\to \psi(2S)$ numerically and
an analytical form  Eq.(\ref{eq3}) is eventually obtained. The three parameters  for the different cases are listed in table \ref{Tab:t2}.
For  $B_c\to \psi(2S)$ transition since  $\psi(2S)$ is a radially  excited state,  two different momentum
distribution amplitudes defined in Eq.(\ref{eq4}) are employed in our numerical calculations.
\begin{table}[!h]
\caption{The decay widths of some modes.}\label{Tab:t3}
\begin{ruledtabular}
\begin{tabular}{ccc}
      &  width (GeV) &  branching ratio   \\\hline
 $B_c\rightarrow J/\psi\pi$  &   $(9.64\pm 2.82)\times 10^{-16} $     &   $(6.64\pm 2.05)\times 10^{-4}$    \\
  $B_c\rightarrow J/\psi K$  &   $(7.66\pm 2.23)\times 10^{-17} $     &   $(5.27\pm 1.62)\times 10^{-5}$    \\
  $B_c\rightarrow J/\psi K^*$  &   $(1.58\pm 0.46)\times 10^{-16} $     &   $(1.09\pm 0.33)\times 10^{-4}$    \\
 $B_c\rightarrow J/\psi D$  &   $(8.02\pm 2.33)\times 10^{-17} $      &    $(5.52\pm 1.69)\times 10^{-5}$   \\
 $B_c\rightarrow J/\psi D^*$  &   $(2.65\pm 0.76)\times 10^{-16} $      &    $(1.82\pm 0.55)\times 10^{-4}$   \\
  $B_c\rightarrow J/\psi D_s$  &   $(1.99\pm 0.58)\times 10^{-15} $      &    $(1.37\pm 0.42)\times 10^{-3}$   \\
 $B_c\rightarrow J/\psi D_s^*$  &   $(5.98\pm 1.72)\times 10^{-15} $      &    $(4.12\pm 1.23)\times 10^{-3}$   \\
 $B_c\rightarrow J/\psi e \bar \nu_e$  &    $(1.67\pm 0.49)\times 10^{-14} $      &   $(1.15\pm 0.36)\%$    \\\hline
  $B_c\rightarrow \psi(2S)\pi$ &  $(4.31\pm 0.42)\times 10^{-16} $      &  $(2.97\pm 0.41)\times 10^{-4}$      \\
    $B_c\rightarrow \psi(2S) K$ &  $(3.34\pm 0.33)\times 10^{-17} $      &  $(2.30\pm 0.32)\times 10^{-5}$      \\
     $B_c\rightarrow \psi(2S) K^*$ &  $(6.37\pm 0.83)\times 10^{-17} $      &  $(4.39\pm 0.71)\times 10^{-5}$      \\
    $B_c\rightarrow \psi(2S)D$ &    $(2.01\pm 0.27)\times 10^{-17} $     &  $(1.38\pm 0.23)\times 10^{-5}$   \\
     $B_c\rightarrow \psi(2S)D^*$ &    $(6.27\pm 1.60)\times 10^{-17} $     &  $(4.32\pm 1.17)\times 10^{-5}$   \\
        $B_c\rightarrow \psi(2S)D_s$ &    $(4.48\pm 0.61)\times 10^{-16} $     &  $(3.08\pm 0.52)\times 10^{-4}$   \\
      $B_c\rightarrow \psi(2S)D_s^*$ &    $(1.29\pm 0.35)\times 10^{-15} $     &  $(8.85\pm 2.54)\times 10^{-4}$   \\
     $B_c\rightarrow \psi(2S) e \bar \nu_e$ &    $(2.73\pm 0.58)\times 10^{-15} $    &    $(1.88\pm 0.44)\times 10^{-3}$   \\\hline
  $B_c\rightarrow \psi_M(2S)\pi$ &  $(2.24\pm 0.19)\times 10^{-16} $     & $(1.54\pm 0.20)\times 10^{-4}$      \\
      $B_c\rightarrow \psi_M(2S)K$ &  $(1.74\pm 0.14)\times 10^{-17} $     & $(1.20\pm 0.15)\times 10^{-5}$      \\
 $B_c\rightarrow \psi_M(2S)K^*$ &  $(3.39\pm 0.24)\times 10^{-17} $     & $(2.33\pm 0.28)\times 10^{-5}$      \\
 $B_c\rightarrow \psi_M(2S)D$ &    $(1.10\pm 0.07)\times 10^{-17} $     &    $(7.57\pm 0.87)\times 10^{-6}$    \\
  $B_c\rightarrow \psi_M(2S)D^*$ &    $(3.55\pm 0.58)\times 10^{-17} $     &    $(2.44\pm 0.47)\times 10^{-5}$    \\
 $B_c\rightarrow \psi_M(2S)D_s$ &    $(2.44\pm 0.14)\times 10^{-16} $     &    $(1.68\pm 0.19)\times 10^{-4}$    \\
$B_c\rightarrow \psi_M(2S)D_s^*$ &    $(7.32\pm 1.29)\times 10^{-16} $     &    $(5.04\pm 1.02)\times 10^{-4}$    \\
     $B_c\rightarrow \psi_M(2S) e \bar \nu_e$ &     $(1.51\pm 0.19)\times 10^{-15} $    &     $(1.04\pm 0.17)\times 10^{-3}$
\end{tabular}
\end{ruledtabular}
\end{table}

With these form factor we calculate the rates for several decay modes. The theoretical predictions are listed in table \ref{Tab:t3} where
the theoretical unertainties are estimated by varying the parameters $m_b$, $m_c$ and $\beta$ within a $10\%$ range.
The predictions of the ratios $\Gamma(B_c\to J/\psi K)/\Gamma(B_c\to J/\psi\pi)$,  $\Gamma(B_c\to J/\psi D_s)/\Gamma(B_c\to J/\psi\pi)$
and $\Gamma(B_c\to J/\psi D_s^*)/\Gamma(B_c\to J/\psi D_s)$ are $0.079\pm 0.033$, $2.06\pm 0.86$ and $3.01\pm 1.23$ respectively which are
consistent with data $0.069\pm0.019\pm0.005$,  $2.9\pm0.57\pm0.24$ and
$2.37\pm0.56\pm0.10$ within 1$\sigma$.

As for the transition $B_c\to \psi(2S)$, by using the two different harmonic oscillator wave functions we obtain
$\Gamma(B_c\to \psi(2S) \pi)/\Gamma(B_c\to J/\psi\pi)=0.45\pm 0.14$ and    $\Gamma(B_c\to \psi_M(2S) \pi)/\Gamma(B_c\to J/\psi\pi)=0.23\pm 0.08$
where the subscript $M$ refers to the modified harmonic oscillator wave function.
The result with the modified harmonic oscillator wave function is obviously closer to the data $0.25\pm 0.068\pm 0.014$ than using the traditional one.
The fact indicates that the modified harmonic oscillator wave functions for radially excited states are more reasonable and applicable.

More theoretical predictions on the channels which have not been yet measured so far are made and
presented in table \ref{Tab:t3}. All the predictions will be tested by future experiments at LHCb or other facilities such as the planned ILC or $Z_0$, Higgs factories etc.
Since the parameter $\beta$ in the wave function of $B_c$ is obtained by an interpolation between the values for $J/\psi$ and $\Upsilon$, it
is not accurate, thus the obtained values of the widths listed in table \ref{Tab:t3} may change for different $\beta$ values, however the ratio between two
widths would  not vary much because the effect caused by the uncertainty of $\beta$ is partly compensated in the ratios.

\section{Summary}
In this paper we calculate the weak decays $B_c\to J/\psi+X$ and
$B_c\to \psi(2S)+X$ within the light-front quark model. The aim of
this work is twofold. The first is to check the validity and
applicability of the modified harmonic oscillator wave function
for radially excited states of heavy quarkonia which we derived in
our earlier work by fitting data of different processes. Secondly,
we further investigate the model parameters which were fixed by
fitting the data of charmonia and bottomonia decays. Namely,  by
comparing our predictions on the rates of several decay modes of
$B_c\to J/\psi+X$ and $B_c\to \psi(2S)+X$  which are the measured
channels, with the available data, the consistency degree confirms
the reasonable range of the model parameters. Then with those
model parameters, we go on predicting the rates for the channels
which have not been measured yet. The predictions will be tested
by the future experiments.

For such $P\to V$ transitions  the form factors were deduced by
several authors \cite{Cheng:2003sm}. With the form factors we
evaluate the rates for semi-leptonic and non-leptonic decays of
$B_c$. Though there is uncertainty for the value of $\beta$ in the
wave function of $B_c$,  the theoretically evaluated ratios
$\Gamma(B_c\to J/\psi K)/\Gamma(B_c\to J/\psi\pi)=0.079\pm 0.033$,
$\Gamma(B_c\to J/\psi D_s)/\Gamma(B_c\to J/\psi\pi)=2.06\pm 0.86$
and $\Gamma(B_c\to J/\psi D_s^*)/\Gamma(B_c\to J/\psi D_s)=3.01\pm
1.23$ are consistent with data within only 1$\sigma$. The rates of
other decays of $B_c\to J/\psi+X$ and $B_c\to \psi(2S)+X$ are also
calculated which will be experimentally  measured soon and by then
we can fix or extract some parameters including the value of
$\beta$ for $B_c$.

In Ref.\cite{Ke:2010x} we suggested a modified harmonic oscillator
wave function for the radially excited states in LFQM. Using these
modified wave functions the obtained decay constants of
$\Upsilon(nS)$ are in good agreement with the data and we also
checked the applicability of these wave functions in the radiative
decays of $\Upsilon(nS)$. In this work we calculate the transition
$B_c\to \psi(2S) \pi$ with the traditional and modified wave
functions for $\psi(2S)$. The theoretical results are quite
different when the two wave functions are employed, as the ratios
are $\Gamma(B_c\to \psi(2S) \pi)/\Gamma(B_c\to
J/\psi\pi)=0.45\pm0.14$ and $\Gamma(B_c\to \psi_M(2S)
\pi)/\Gamma(B_c\to J/\psi\pi)=0.23\pm0.08$ and the result using
the modified wave function is closer to the data
$0.25\pm0.068\pm0.014\pm0.006$. Namely, our numerical results
which are satisfactorily consistent with data of $B_c\to
\psi(2S)+X$, indicate that the modified wave function works better
than the traditional one not only for the radially excited
bottomonia, but also for radially excited charmonia. The
consistency degree of other predictions for $B_c\to \psi(2S)+X$
with the future experimental data will provide further test to the
modified wave function.

\section*{Acknowledgement}

This work is supported by the National Natural Science Foundation
of China (NNSFC) under the contract No. 11075079 and No. 11005079;
the Special Grant for the Ph.D. program of Ministry of Eduction of
P.R. China No. 20100032120065.

\appendix

\section{}

%\appendix

\end{document}